\documentclass[aps,showpacs,prl,twocolumn]{revtex4}
\usepackage{amsmath,amsfonts}
\usepackage{graphicx}

\def\l{\left}
\def\r{\right}
\def\be{\begin{equation}}
\def\ee{\end{equation}}
\def\bea{\begin{eqnarray}}
\def\eea{\end{eqnarray}}

\newcommand{\ket}[1]{\mbox{$\left| #1 \right\rangle$}}

\newcommand{\eq}[1]{Eq.~(\ref{eq:#1})}

\def\cG{{\cal G}}

\def\cO{{\cal O}}
\def\cP{{\cal P}}
\def\cQ{{\cal Q}}
\def\cS{{\cal S}}
\def\cU{{\cal U}}
\def\bbC{\mathbb{C}}
\def\bU{{{\bf U}}}
\def\>{\rangle}
\def\<{\langle}
\def\lb{\left[}
\def\rb{\right]}

\def\lbL{\lb\rule{0pt}{2.4ex}}

\begin{document}

\title{Efficient Quantum Circuits for Schur and Clebsch-Gordan Transforms}
\author{Dave Bacon$^\dagger$, Isaac L. Chuang$^\ddagger$, and
        Aram W. Harrow$^\ddagger$}
\affiliation{$^\dagger$Institute for Quantum Information \&
Department
of Physics \\
California Institute of  Technology, Pasadena, CA 91125 USA \\
$^\ddagger$Center for Bits and Atoms \& Department of Physics \\
 Massachusetts Institute of Technology,
Cambridge, MA 02139 USA}
\date{\today}
\begin{abstract}
The Schur basis on $n$ $d$-dimensional quantum systems is a
generalization of the total angular momentum basis that is useful
for exploiting symmetry under permutations or collective unitary
rotations.  We present efficient (size
$\text{poly}(n,d,\log(1/\epsilon))$ for accuracy $\epsilon$) quantum
circuits for the {\em Schur transform}, which is the change of basis
between the computational and the Schur bases. These circuits are
based on efficient circuits for the Clebsch-Gordan transformation.
We also present an efficient circuit for a limited version of the
Schur transform in which one needs only to project onto different
Schur subspaces.  This second circuit is based on a generalization
of phase estimation to any nonabelian finite group for which there
exists a fast quantum Fourier transform.
\end{abstract}
\maketitle

A key component of quantum algorithms is their ability to reveal
information stored in non-local degrees of freedom.  In particular,
one of the most important building blocks known is the quantum
Fourier transform (QFT)\cite{Shor:96a}, an efficient circuit
construction for conversion between discrete position and momentum
bases.  The QFT converts a vector of $2^n$ amplitudes in $O(n^2)$
steps, in contrast to the $O(n2^n)$ which would be required
classically.

Another elementary basis change important in quantum physics is
between independent local states and those of definite total angular
momentum.  When two identical spins interact with a global
excitation, due to their permutation symmetry they appear as a
singlet or a triplet to the external interaction.  Such states of
definite permutation symmetry can naturally hold entanglement, a
physical resource central to quantum information.

The basis transformation defined by permutation symmetry is also
central to a plethora of quantum information protocols.  These
include methods to estimate the spectrum of a density
operator\cite{Keyl:01a}, achieve optimal quantum hypothesis
testing\cite{Hayashi:02d}, perform universal quantum source
coding\cite{Hayashi:02b}, concentrate entanglement in a
distortion-free manner\cite{Hayashi:02a}, create decoherence-free
states\cite{Knill:00a}, and communicate without a shared reference
frame\cite{Bartlett:03a}.  This deep connection is a natural
consequence of the underlying states and random variables being
independent and identically distributed.  However, unlike the QFT,
the complexity of this transform has been unknown, rendering
protocols which use it non-constructive.  Also, to be useful in
quantum algorithms, this transform must be {\em efficient}, that is,
constructible for an $n$ spin system using a quantum circuit of
depth polynomial (versus exponential) in $n$.

Here, we resolve this problem by giving an efficient construction
for performing a transformation between local and total angular
momentum descriptions of a set of $n$ $d$-dimensional systems ($n$
``qudits''), for arbitrary $n$ and $d$.  This is achieved using a
quantum circuit of size $\text{poly}(n,d,\log(1/\epsilon))$ for
accuracy $\epsilon$. We believe that this basis change, which we
call the {\em Schur transform}, is important not only for quantum
information, but also as a new building block for future quantum
algorithms.  This is demonstrated by a close connection between the
Schur transform and a generalization of the quantum phase estimation
algorithm\cite{Cleve:98a} for non-abelian groups, as discussed
below.

{\em The Schur Transform ---} Consider a system of $n$ qudits, each
with a standard local (``computational'') basis $|i\>$, $i=1\dots
d$. The Schur transform relates transforms on the system performed
by local $d$-dimensional unitary operations to those performed by
permutation of the qudits.  Recall that the symmetric group
${\mathcal S}_n$ is the group of all permutations of $n$ objects.
This group is naturally represented in our system by \be
    {\bf P}(\pi) |i_1 i_2 \cdots i_n\> = |i_{\pi^{-1}(1)}
        i_{\pi^{-1}(2)} \cdots i_{\pi^{-1}(n)}\>
\,, \label{eq:pdef} \ee where $\pi \in {\mathcal S}_n$ is a
permutation and $|i_1 i_2\ldots \>$ is shorthand for
$\ket{i_1}\otimes\ket{i_2}\otimes\ldots$.  Let ${\mathcal U}_d$
denote the group of $d\times d$ unitary operators. This group is
naturally represented in our system by \be
   {\bf Q}(U)|i_1 i_2 \cdots i_n\>
    = U |i_1\> \otimes U|i_2\> \otimes \cdots \otimes U|i_n\>
\,, \label{eq:qdef} \ee where $U \in {\mathcal U}_d$.

The Schur transform is based on Schur duality, a well
known\cite{Goodman:98a} and powerful way to relate the
representation theory of ${\bf P}(\pi)$ and ${\bf Q}(U)$. For
example, consider the case of two qubits ($n=2$, $d=2$).  The
two-qubit Hilbert space $(\bbC^2)^{\otimes 2}$ decomposes under
${\bf Q}$ into a one-dimensional spin-$0$ singlet space and a
three-dimensional spin-$1$ triplet space.  Both of these are
irreducible representations (irreps) of ${\mathcal U}_2$, but they
also happen to be irreps of ${\mathcal S}_2$.  The singlet state
changes sign under permutation of the two spins, and the triplet
states are invariant under permutation.  These correspond to the
sign ${\cal P}_{\text{sign}}$ and the trivial ${\cal
P}_{\text{trivial}}$ irreps of $\cS_2$, and thus we can write
$(\bbC^2)^{\otimes 2} \cong ( {\cal Q}_1 \otimes {\cal
P}_{\text{trivial}}) \oplus ({\cal Q}_0 \otimes {\cal
P}_{\text{sign}})$, where ${\cal Q}_J$ is the spin-$J$ irrep of
${\cal U}_2$.

This unusual coincidence between the two representations exists for
an arbitrary number of qudits, becoming quite non-trivial for larger
$n$ and $d$.  For example, the Hilbert space of three qubits ($n=3,
d=2$) decomposes into $({\cal Q}_{3/2} \otimes {\cal
P}_{\text{trivial}} ) \oplus ({\cal Q}_{1/2} \otimes {\cal P}_{2,1}
)$, where ${\cal P}_{2,1}$ denotes a particular two-dimensional
mixed symmetry irrep of $S_3$.  In terms of the original (local)
basis the ${\cal Q}_{1/2} \otimes {\cal P}_{2,1}$ space contains two
spin-$1/2$ objects, one spanned by $\ket{110} + \omega\ket{011} +
\omega^*\ket{101}$ (suppressing normalization) and $\ket{001} +
\omega\ket{100} + \omega^*\ket{010}$, and the other obtained by
replacing $\omega = e^{2\pi i/3}$ with $\omega^*$.  These two spaces
correspond to the two mixed symmetries in ${\cal P}_{2,1}$.

The general theorem of Schur duality states that for any (integer)
$d$ and $n$, \be
    \l(\bbC^d\r)^{\otimes n} \cong \bigoplus_{\lambda\in {\rm Part}[n,d]}
        \cQ_\lambda \otimes \cP_\lambda
\,, \label{eq:schur-spaces} \ee
where $\lambda$ is chosen from the set of possible partitions of $n$
into $\leq d$ parts.  Thus, there exists a basis for
$(\bbC^d)^{\otimes n}$ with states
$|\lambda,q_\lambda,p_\lambda\>_{\rm Sch}$ where $\lambda$ labels
the subspaces $\cQ_\lambda \otimes \cP_\lambda$ and $q_\lambda\in
\cQ_\lambda$ and $p_\lambda \in \cP_\lambda$ label bases for
$\cQ_\lambda$ and $\cP_\lambda$ respectively. We may represent these
states of generalized definite total angular momentum and
permutation multiplicity by vectors $|\lambda,q,p\>$ in the
computational basis, with bit strings $\lambda$, $q$, and $p$.  Note
that  ${\rm dim}(\cQ_\lambda)$ and ${\rm dim}(\cP_\lambda)$ vary
with $\lambda$, and so $|q\>$ and $|p\>$ are padded; this requires
only constant spatial overhead.

Just as in the examples above, the Schur basis states
$|\lambda,q_\lambda,p_\lambda\>_{\rm Sch}$ are superpositions of the
$n$ qudit computational basis states $|i_1 i_2 \ldots i_n\>$,

\begin{equation}
    |\lambda,q_\lambda,p_\lambda\>_{\rm Sch}
= \sum_{i_1,\ldots,i_n}
    \lbL{ {\bf U}_{\rm Sch} }\rb^{\lambda,q,p}_{i_1,i_2,\dots,i_n}
    |i_1 i_2 \cdots i_n\>
\,.
\end{equation}
By the isomorphism of Eq.(\ref{eq:schur-spaces}), this defines a
unitary transformation ${\bf U}_{\rm Sch}$ (with matrix elements as
given), the Schur transform we desire.  ${\bf U}_{\rm Sch}$ maps the
computational basis to the $|\lambda,q,p\>$ representation of the
Schur basis.

Applying the Schur transform extracts $\lambda$, $q$, and $p$ values
for a given state, allowing the values be manipulated like any other
quantum data. For example, since maximally entangled states are
invariant under permutation, if $|\psi\>_{AB}$ is a bipartite
partially entangled state, then a universal scheme for entanglement
concentration from $|\psi\>^{\otimes n}$ is given by both parties
performing the Schur transform ${\bf U}_{\rm Sch}$, measuring
$|\lambda\>$, discarding $|q\>$ and noting that $|p\>$ contains a
maximally entangled state of Schmidt rank
$\dim(\cP_\lambda)$\cite{Hayashi:02a}.  It is then straightforward
to map $\ket{p}$ reversibly onto the integers
$\{1,\ldots,\dim(P_\lambda)\}$ to obtain the state
$\frac{1}{\sqrt{d}}\sum_{i=1}^d\ket{ii}_{AB}$ for
$d=\dim(\cP_\lambda)$ \cite{Bacon:05a}.

The defining property of ${\bf U}_{\rm Sch}$ is that it reduces the
action of ${\bf Q}$ and ${\bf P}$ into irreps. For any $\pi\in S_n$
and any $U\in\cU_d$, ${\bf P}(\pi)$ and ${\bf Q}(U)$ commute, so we
can express both reductions at once as
\be
    {\bf U}_{\rm Sch}{\bf Q}(U){\bf P}(\pi){\bf U}_{\rm Sch}^\dag
    =
    \sum_{\lambda\in{\rm Part}(d,n)}
        |\lambda\>\<\lambda| \otimes
        {\bf q}_\lambda(U) \otimes
        {\bf p}_\lambda(\pi)
\,, \ee
where ${\bf q}_\lambda$ and ${\bf p}_\lambda$ are irreps of
${\mathcal U}_d$ and ${\mathcal S}_n$ respectively.

{\em Quantum Circuit for the Schur Transform ---} We construct a
quantum circuit\cite{Nielsen:00a} for ${\bf U}_{\rm Sch}$ in two
stages, first for $d=2$, then generalizing to $d>2$.  Each of these
constructions follows an iterative structure, in which the Schur
transform on $n$ qudits is realized using $O(n)$ elementary steps,
each of which adds a single qudit to an existing Schur state of the
form $|\lambda,q,p\>$.

For $d=2$, this elementary step is familiar from basic quantum
mechanics, because it involves simple addition of angular momentum,
following the prescription for calculation of the Clebsch-Gordan
(CG) coefficients\cite{Sakurai:94a}.  In this case, $\lambda$ and
$q$ can be conveniently denoted by half-integers $J$ and $m$ (with
$|m|\leq J \leq n/2$) which give the total angular momentum and the
$z$-component of angular momentum respectively.  And in terms of
$J$, the CG transform takes as input $|J,m\>$ and a single spin
$|s\>$, and outputs a linear combination of the states $|J\pm 1/2,
m\pm 1/2\>$. The amplitudes of the linear combination are readily
computed using the usual ladder operators for raising and lowering
angular momenta. In addition, however, we must distinguish between
multiple distinct pathways which add up to give the same total $J$,
as demonstrated by the three qubit example above.  In fact, it is
the permutation symmetry of these pathways which give rise to
$\cP_J$, and thus we track the pathway with another output label
$p=J'-J$.

\begin{figure}[htbp]
\begin{center}
\includegraphics[scale=1.00]{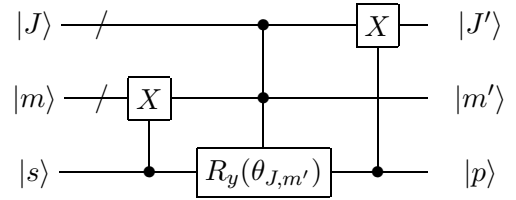}
\end{center}
\vspace*{-2ex} \caption{Quantum circuit implementing $\bU_{\rm CG}$
to convert between the $|J,m\>|s\>$ and $|J',m',p\>$ bases, for the
$d=2$ (qubit) case. Following standard
conventions\cite{Nielsen:00a}, time goes from left to right, the
$|J\>$ and $|m\>$ wires hold multiple qubits, and $|s\>$ is one
qubit.  The controlled $X$ operation ${\bf C}_{X}$ adds the control
to the target qubits, i.e. ${\bf C}_{X} |s\>|m\>=|s\>|m+s\>$. The
doubly controlled $R_y(\theta_{J,m^\prime})$ gate implements the
rotation given by Eq.~(\ref{eq:rot}) using the $J$ and $m^\prime$
qubits.} \label{fig:littlecircuit}
\end{figure}

Putting this together, we can define an elementary Clebsch-Gordan
transform step ${\bf U}_{\rm CG}$ as a rotation between two specific
basis states,
\begin{eqnarray}
  && \left[\begin{array}{c}
            |J_{-}^\prime,m^\prime,p=-\frac{1}{2} \> \\
        |J_{+}^\prime,m^\prime,p=+\frac{1}{2}\>
     \end{array}
\right]
\nonumber \\
&&~ =
    \left[ \begin{array}{cc} \cos{\theta_{J,m'}} & -\sin{\theta_{J,m'}} \\
                  \sin{\theta_{J,m'}} & \cos{\theta_{J,m'}}
              \end{array} \right]
          \left[\begin{array}{c}
               |J,m_{+}\>|s=-\frac{1}{2}\> \\
           |J,m_{-}\>|s=+\frac{1}{2}\>
          \end{array} \right]
\,, \label{eq:rot}
\end{eqnarray}
where $J'_{\pm} = J\pm 1/2$, $m_{\pm} = m'\pm{1}/{2}$, and
$\cos{\theta_{J,m'}} =\sqrt{ \frac{J+ m^\prime+1/2}{2J+1}}$.  ${\bf
U}_{\rm CG}$ can be realized with three gates in a quantum circuit,
as shown in Fig.~\ref{fig:littlecircuit}, using as one gate a
controlled rotation about $\hat y$ by angle $\theta_{J,m^\prime}$.
This angle is computed using usual quantum and reversible circuit
techniques\cite{Nielsen:00a} with error $\epsilon$, using ${\rm
poly}(\log(1/\epsilon))$ standard circuit elements.

\begin{figure}[htbp]
\begin{center}
\includegraphics[width=0.48\textwidth]{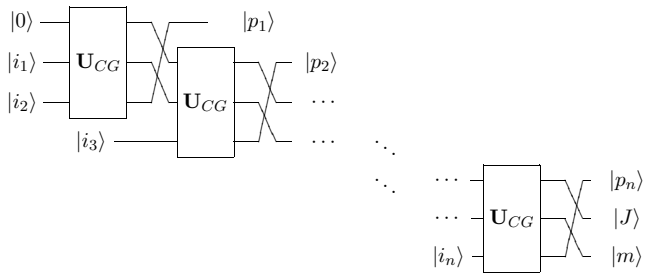}
\end{center}
\vspace*{-3.5ex} \caption{Quantum circuit for the Schur
transformation $\bU_{\rm Sch}$, transforming between $|i_1 i_2
\cdots i_n\>$ and $|J,m,p\>$.} \label{fig:schur-circuit}
\end{figure}

The full Schur transform is implemented by cascading $\bU_{\rm CG}$
as shown in Fig.~\ref{fig:schur-circuit}. The complexity of this
circuit is thus $\cO(n\cdot\text{poly}\log(1/\epsilon))$.  In the
circuit, $p_1 p_2 \cdots p_n$ is an encoding in Young's orthogonal
basis\cite{Beals:97a} for $p$, the label capturing the permutation
symmetry of the state.  This basis is an example of a subgroup
adapted basis, a useful type of basis used by Beals in the fast
quantum Fourier transform over the symmetric group\cite{Beals:97a}.
For $d=2$, the total angular momentum of the state, $J$, gives the
partition $\lambda$.

Construction of the Schur transform for $d>2$ follows the same ideas
as for $d=2$, but is complicated by the necessary notation; the
principle challenge is showing that the elementary $\bU_{\rm CG}$
steps for $d>2$ can be computed in ${\rm poly}(d)$ steps. $\bU_{\rm
Sch}$ is constructed as a cascade of $O(n)$ $\bU_{\rm CG}$
transforms, just as for $d=2$.  Each $\bU_{\rm CG}$ combines an
arbitrary irrep of $\cU_d$, a multi-qudit state $|\lambda,q\>$, with
a single qudit state $|i_k\>$, to obtain a multi-qudit superposition
of new irreps of $\cU_d$, $|\lambda',q'\>$.  Simultaneously, the
permutation labels $|p\>$ are constructed; equivalently we could
save the values of $\lambda$ that we generate in each step.
$\bU_{\rm CG}$ can be computed efficiently because of a recursive
relationship between $\bU_{\rm CG}$ for ${\mathcal U}_d \times
{\mathcal U}_d$ and that of ${\mathcal U}_{d-1} \times {\mathcal
U}_{d-1}$ in terms of reduced Wigner coefficients\cite{Louck:70a}.
Crucially, there is an efficient classical algorithm for the
computation of the reduced Wigner coefficients\cite{Biedenharn:68a}
needed for $\bU_{\rm CG}$. Specific details of this calculation are
given elsewhere\cite{Bacon:05a}. The complexity of the full Schur
transform is thus found to be polynomial in $n$, $d$, and
$\log(\epsilon^{-1})$.

{\em Circuit for Schur Basis Measurement ---}
Unitary transforms enable measurements in alternate bases; for such
applications, simpler quantum circuits can be employed, as we now
demonstrate for measuring $\lambda$ in the Schur basis.
Fascinatingly, the quantum circuit we construct utilizes the same
structure as that of the quantum factoring algorithm\cite{Shor:96a},
and its generalizations to non-abelian
groups\cite{Moore:03a,Beals:97a,Kitaev:97a,Klappenecker:03a}.

\begin{figure}[htbp]
\begin{center}
\includegraphics[scale=1.00]{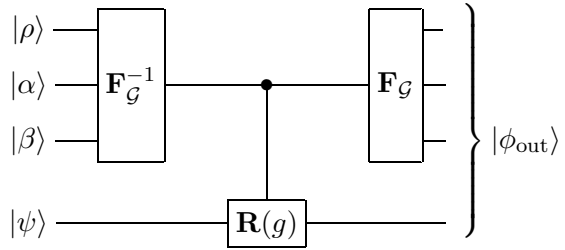}
\end{center}
\vspace*{-3.5ex} \caption{Quantum circuit used in measurement of an
irrep of a group $\cG$.} \label{fig:genkit}
\end{figure}

Measurement of $\lambda$ projects a state onto the different spaces
$\cQ_\lambda \otimes \cP_\lambda$, as defined in \eq{schur-spaces}.
For any representation of a group for which there exists a quantum
Fourier transform, we give a circuit for performing this projective
measurement onto irreps in Fig.~\ref{fig:genkit}; when specialized
to the symmetric group and the representation ${\bf P}(\pi)$, this
can be used to measure $\lambda$.

The circuit uses the following gate elements.  Let ${\mathcal G}$ be
an arbitrary finite group over which there exists an efficient
quantum circuit for the quantum Fourier transform \cite{Moore:03a},
${\bf F}_{\mathcal G}$.  Define $\hat{\mathcal G}$ to be a
representative set of inequivalent irreps of ${\mathcal G}$ and let
$d_\rho$ denote the dimension of the irrep $\rho\in \hat{\mathcal
G}$.  ${\bf F}_{\mathcal G}$ is then a unitary transform from a
space spanned by group elements, $|g\>$, $g \in {\mathcal G}$ to a
space spanned by irrep labels and row and column indices, $|\rho$,
$\alpha$, $\beta\>$, $\rho \in \hat{\mathcal G}$, and
$\alpha,\beta=1\dots d_\rho$.  Specifically,
\begin{equation}
    {\bf F}_{\mathcal G} = \sum_{\rho \in \hat{\mathcal G}}
    \sum_{\alpha,\beta=1}^{d_\rho} \sum_{g \in {\mathcal G}}
    \sqrt{\frac{d_\rho}{|{\mathcal G}|}}
    \lbL{{\bf r}_\rho(g)}\rb_{\alpha \beta}
    |\rho,\alpha,\beta\> \<g|
\end{equation}
where $\lbL{{\bf r}_\rho(g)}\rb_{\alpha \beta}$ is the entry of the
$\alpha$th row and $\beta$th column of the irrep $\rho$ evaluated at
$g \in {\mathcal G}$. Finally, let ${\bf R}(g)$, $g\in{\mathcal G}$
be a generic representation of this group ${\mathcal G}$ and suppose
we can efficiently enact the controlled ${\bf R}(g)$ operation,
${\bf C}_{\bf R}=\sum_{g \in {\mathcal G}} |g\> \<g| \otimes {\bf
R}(g)$.  ${\bf F}_{\mathcal G}$, its inverse, and ${\bf C}_{\bf R}$
are the circuit elements used to construct the circuit in
Fig.~\ref{fig:genkit}.

This circuit is employed in the following manner, to measure an
irrep of ${\mathcal G}$.  Consider the action of the circuit in
Fig.~\ref{fig:genkit} when a generic input state $|\psi\>$ is fed
into the space upon which the generic representation of the group
${\bf R}(g)$ acts. Since ${\bf R}(g)$ is a generic representation of
the group ${\mathcal G}$, the space upon which this representation
acts can be decomposed into different irreps of ${\mathcal G}$.  Let
the multiplicity of the $\mu$th irrep ($\mu \in \hat{\mathcal G}$)
in ${\bf R}(g)$ be $n_\mu$.  It is possible that an irrep $\mu$ does
not appear at all, in which case $n_\mu=0$. Then
\begin{equation}
|\psi\>=\sum_{\mu \in \hat{\mathcal G}} \sum_{j=1}^{n_\mu}
\sum_{k=1}^{d_\mu} c_{\mu,j,k}|\mu,j,k\>_{\rm R}
\end{equation}
meaning we can expand the generic input state $|\psi\>$ over a basis
labeled by the irrep labels $\mu$, multiplicity labels for the irrep
$j$, and a label for the spaces upon which the irrep acts $k$. Since
this basis fully reduces ${\bf R}(g)$,
\begin{equation}
{\bf R}(g)|\psi\>=\sum_{\mu \in \hat{\mathcal G}}
    \sum_{j=1}^{n_\mu} \sum_{k=1}^{d_\mu} c_{\mu,j,k}|\mu,j\> {\bf
    r}_\mu(g)|k\>_{\rm R}
\,. \label{eq:irrepact}
\end{equation}
The output of the circuit in Fig.~\ref{fig:genkit} is thus
\begin{equation}
    |\phi_{\rm out}\> =
    \sum_{g \in {\mathcal G}}
    \sum_{\rho,\alpha,\beta} \frac{\sqrt{d_\rho}}{|{\mathcal G}|}
    \lbL{ {\bf r}_\rho(g) }\rb_{\alpha,\beta}
    |\rho,\alpha,\beta\> \otimes {\bf R}(g) |\psi\>  
\,. \label{eq:psiout}
\end{equation}
Using the reducible action of ${\bf R}(g)$ on $|\psi\>$ given by
Eq.(\ref{eq:irrepact}) along with the orthogonality relationships
for irreps\cite{Goodman:98a}, Eq.(\ref{eq:psiout}) can be
reexpressed as
\begin{equation}
    |\phi_{\rm out}\> =
    \sum_{\mu \in \hat{\mathcal G}} \sum_{j=1}^{n_\mu}
    \sum_{\alpha,\beta=1}^{d_\alpha} \frac{c_{\mu,j,\alpha}}{
    \sqrt{d_\mu}} |\mu,\alpha,\beta \> \otimes
    |\mu,j,\beta\>_{\rm R}
\,.
\end{equation}

The output $|\phi_{\rm out}\>$ has multiple interesting properties
which we can now exploit.  Measuring the first register (the irrep
label index) produces outcome $\mu$ with probability
$\sum_{j=1}^{n_\mu} \sum_{k=1}^{d_\mu} |c_{\mu,j,k}|^2$.  This is
exactly the probability we would obtain if we were to measure the
irrep label of $|\psi\>$.  Remarkably, this is achieved independent
of the basis in which ${\bf C}_{\bf R}$ is implemented.  For the
Schur basis, taking ${\mathcal G}={\mathcal S}_n$ and letting ${\bf
R}={\bf P}$ denote the natural representation of the symmetric group
given by Eq.(\ref{eq:pdef}) so that $\mu=\lambda$, the measurement
circuit gives exactly the probabilities we would obtain if we were
to compute $\bU_{\rm Sch}|\psi\>$ then measure $\lambda$.  This
works for the Schur transform over not just qubits, but also over
qudits. Furthermore, for large qudit dimension $d$, the new circuit
is exponentially faster than the full Schur transform $\bU_{\rm
Sch}$ followed by measurement; in terms of $d$, $\bU_{\rm Sch}$ has
size $\text{poly}(d)$, while this measurement circuit is
$\text{poly}\log(d)$.

The circuit in Fig.~\ref{fig:genkit} is a generalization of the
phase estimation circuit introduced by Kitaev\cite{Kitaev:95a},
which uses a QFT over an abelian group (for phase estimation, a
$U(1)$ phase is approximated by the cyclic group ${\mathcal C}_N$).
In contrast, our circuit is applied to non-abelian groups; and in
addition to allowing measurement of group irreps, this circuit also
allows operations to be performed in the irreducible basis of ${\bf
R}(g)$, for arbitrary groups ${\mathcal G}$.  This cannot be
accomplished directly with $\bU_{\rm Sch}$, which is specialized to
the permutation and unitary groups.

{\em Conclusion---}We have shown how to efficiently perform the
Schur transform and shown how a generalization of Kitaev's circuit
can be used to perform a useful group representation transformation.
Whenever we have identical copies of a quantum system or are
averaging over the diagonal action of the unitary group this first
transform is often essential.  Whenever the symmetry of a quantum
state is described by a finite group, the generalized Kitaev
transform comes into play.  One of the most fundamental problems in
quantum computation is the search for new quantum algorithms.  In
this respect, there are few unitary transforms which have both an
efficient quantum circuit and interpretations which might allow
these transforms to be useful in an algorithm. We are hopeful that
our circuits will be useful exactly because they have such clear
group representation theory interpretations.

{\em Acknowledgments---}This work was partially funded by the NSF
Center for Bits and Atoms contract CCR-0122419 and the NSF Institute
for Quantum Information under grant number EIA-0086048; AWH was
supported by the NSA and ARDA under ARO contract DAAD19-01-1-06. We
thank Nolan Wallach for useful discussions.

\end{document}